\documentclass[10pt]{article}
\usepackage{amsmath}
\usepackage{amssymb}

\usepackage{subfigure}
\usepackage{psfrag}
\usepackage{epsfig}
\usepackage{graphicx}
\usepackage{tikz}

\newcount\xrefpos \xrefpos=0
\newcount\yrefpos \yrefpos=0
\newcount\xput \xput=0
\newcount\yput \yput=0

\def\refpos#1 #2 #3{\global\xrefpos=#1 \global\yrefpos=#2
                         \rlap{$\smash{#3}$}}
\def\put #1 #2 #3{\xput=#1 \yput=#2
                  \advance\xput by -\xrefpos
                  \advance\yput by -\yrefpos
                  \rlap{\kern\the\xput truebp
                        \vbox to 0pt{\vss\hbox{$\displaystyle #3$}
                        \kern\the\yput truebp}}}
\def\beginlabels\refpos#1\endlabels{\hbox{$\refpos#1$}}

\usepackage{amsfonts,amssymb}
\newcommand{\bea}{\begin{eqnarray}}
\newcommand{\eea}{\end{eqnarray}}
\newcommand{\eref}[1]{Eq.~(\ref{#1})}
\newcommand{\nn}{\nonumber}
\usepackage{mciteplus}

\begin{document}

 \begin{center}
  {\Large \bf Hard-gapped Holographic Superconductors}

\vspace{3mm}

Pallab Basu$^{a,}$\footnote{email: pallab@phas.ubc.ca} , Jianyang He $^{a}$\footnote{email: jyhe@phas.ubc.ca}, Anindya Mukherjee$^{a,}$\footnote{anindya@phas.ubc.ca} and Hsien-Hang Shieh$^{a,b}$\footnote{shieh@phas.ubc.ca}

\bigskip\medskip
\centerline{$^a$\it Department of Physics and Astronomy}
\smallskip\centerline{\it University of British Columbia}
\smallskip\centerline{\it Vancouver, BC V6T 1Z1, Canada}

\bigskip \centerline{$^b$\it Perimeter Institute for theoretical Physics}
\smallskip\centerline{\it 31 Caroline Street North}
\smallskip\centerline{\it Waterloo, Ontario }
\smallskip\centerline{\it Canada N2L 2Y5}
 \end{center}

\abstract{In this work we discuss the zero temperature limit of a ``p-wave'' holographic superconductor. The bulk description consists of a non-Abelian $SU(2)$ gauge fields minimally coupled to gravity. We numerically construct the zero temperature solution which is the gravity dual of the superconducting ground state of the ``p-wave'' holographic superconductors. The solution is a smooth soliton with zero horizon size and shows an emergent conformal symmetry in the IR. We found the expected superconducting behavior. Using the near horizon analysis we show that the system has a ``hard gap'' for the relevant gauge field fluctuations. At zero temperature the real part of the conductivity is zero for an excitation frequency less than the gap frequency. This is in contrast with what has been observed in similar scalar-gravity-gauge systems (holographic superconductors). We also discuss the low but finite temperature behavior of our solution. }

\section{Introduction}
In recent years string theory or, more specifically, gauge-gravity duality has seen interesting applications in the field of condensed matter physics. One of the earliest such applications is the discovery of a superconductor-like phase transition in AdS with a Reissner-Nordstr\"{o}m black hole and a charged scalar field minimally coupled to a $U(1)$ gauge field \cite{Gubser:2008px,Hartnoll:2008vx}.

Earlier works on this system at low (but non zero) temperatures indicated the presence of a gap in the spectrum, with an exponentially decaying population of normal carriers following the Boltzmann distribution. In various recent works \cite{Horowitz:2009ij,Gubser:2009cg,Gubser:2009gp, Gubser:2008pf, Gauntlett:2009dn, Gubser:2008wz, Konoplya:2009hv}, authors have addressed the problem in the zero temperature limit. More precisely, they consider a charged scalar field and gauge fields in an $AdS_4$ background at zero temperature. The solution turns out to be a solitonic solution with a zero sized horizon. The authors find that the effective potential for small gauge field fluctuations vanishes near the black hole horizon. This implies that the normal component of the A.C. conductivity never vanishes, even at zero temperature, which in turn indicates that the superconductor is gapless. This is a new and interesting result different from what would have been naively expected.

In this paper we consider non-Abelian gauge fields in the background of an extremal $AdS_4$ black hole. The setup is quite similar to so-called ``p-wave'' holographic superconductors \cite{Gubser:2008wv} and known to have a superconducting transition.\footnote{For other works in non-Abelian holographic superconductor look at \cite{Gubser:2008zu, Basu:2008bh,Roberts:2008ns,Ammon:2008fc,Peeters:2009sr, Franco:2009yz,Herzog:2009ci,Zeng:2009dv}.}.  We work at the zero temperature and found a fully gravity back-reacted solitonic solution to this system similar to \cite{Horowitz:2009ij}. Then we consider the frequency dependent conductivity by considering certain vector fluctuations on this background. This is an anisotropic system which shows different conductivity at different directions. We find the expected superconducting behavior for relevant currents. Interestingly we find that the corresponding effective potential does not vanish at the horizon (it goes to zero near the boundary). Following the argument in \cite{Horowitz:2009ij}, we thus arrive at the conclusion that the holographic non-Abelian superconductor does have a finite gap for the relevant gauge field fluctuations.

The plan of the paper is as follows, in  section \ref{sec:setup} we discuss the general setup and Einstein's equation for a non-Abelian gauge field coupled to gravity. In section \ref{sec:RN} we discuss the instability of the RN black hole solution. In section \ref{sec:zero} we construct the zero temperature solitonic solution which is the superconducting ground state of the theory. We also discuss issues related to conductivity and show the existence of a hard gap. In section \ref{sec:finite} we propose a possible form of the solution at non-zero temperature using the method of matched asymptotic expansion.

\section{Setup}
\label{sec:setup}
The Einstein-YM action for a non-Abelian gauge field with a negative cosmological constant is given by \cite{Gubser:2008zu},
\begin{eqnarray}
{\cal L} =\int d^4x\sqrt{-g}\left({\mathcal
{R}}+\frac{6}{l^2}-\frac{1}{4}F_a^{\mu\nu}F_{\mu\nu}^a\right),
\end{eqnarray}
where $F_{\mu\nu}$ is the field strength of an $SU(2)$ gauge field.

To tally with our plan of considering anisotropy in the spatial direction, we choose the following ansatz for our metric
\begin{eqnarray}
ds^2=-g(r)e^{-\chi(r)}dt^2+\frac{dr^2}{g(r)}+r^2\Big(c(r)^2 dx^2+dy^2\Big),
\end{eqnarray}
and the gauge fields \footnote{Due to a repulsive term coming from the non-Abelian interactions, it is expected that a isotropic ansatz will have a quartic instability and would possibly have more free energy than the anisotropic ones \cite{Gubser:2008wv,Basu:2008bh}. We left the detailed discussion of these issues for a future study.},
\begin{eqnarray}
A=A(r)\tau^3 dt+B(r)\tau^1 dx.
\end{eqnarray}

It is straightforward to find the non-zero components of the field strength
\begin{eqnarray}
F_{\mu\nu}^a=\partial_{\mu}A_{\nu}^a-\partial_{\nu}A_{\mu}^a+i q[A_{\mu},A_{\nu}]^a
\end{eqnarray}
are
\begin{eqnarray}
F_{rx}^1=-F_{xr}^1=B'(r),~~~ F_{rt}^3=-F_{tr}^3=A'(r),~~~F_{xt}^2=-F_{tx}^2=qAB.
\end{eqnarray}
The energy momentum tensor is
\begin{eqnarray}
T_{\mu\nu}=-\frac{1}{\sqrt{-g}}\frac{\delta S_{matter}}{\delta g^{\mu\nu}}
=-\frac{1}{2}g_{\mu\nu}\left(\frac{1}{4}F^2\right)+\frac{1}{2}g^{\alpha\beta}F_{\mu\alpha}^aF_{\nu\beta}^a,
\end{eqnarray}
and the non-zero components of the energy momentum tensor are,
\begin{eqnarray}
T_{tt}&=& \frac{1}{4}g A'^2+\frac{g^2}{4r^2 c^2}e^{-\chi}B'^2+\frac{1}{4r^2 c^2}(qAB)^2, \nonumber \\
T_{rr}&=& -\frac{e^{\chi}}{4g c^2}A'^2+\frac{1}{4r^2 c^2}B'^2+\frac{e^{\chi}}{4g^2r^2 c^2}(qAB)^2, \nonumber\\
T_{xx}&=& \frac{1}{4}e^{\chi}r^2 c^2 A'^2+\frac{g}{4}B'^2-e^{\chi}\frac{(qAB)^2}{4g}, \nonumber\\
T_{yy}&=& \frac{1}{4}e^{\chi}r^2 A'^2-\frac{g}{4c^2}B'^2+e^{\chi}\frac{(qAB)^2}{4g c^2}.
\end{eqnarray}
It is to be noted that here $T_{xx} \neq T_{yy}$ due our anisotropic ansatz\footnote{HSH thanks Matthew Roberts and Gary Horowitz for discussions on this point.}.


The Maxwell's equations of $A(r),B(r)$ are
\begin{eqnarray}
A_t^3\longrightarrow& A''+A'\left(\frac{2}{r}+\frac{\chi'}{2}+\frac{c'}{c}\right)-\frac{q^2B^2}{r^2gc}A=0, \nonumber\\
A_x^1\longrightarrow& B''+B'\left(\frac{g'}{g}-\frac{\chi'}{2}-\frac{c'}{c}\right)+\frac{e^{\chi}q^2A^2}{g^2}B=0.
\label{maineq}
\end{eqnarray}
The potential term for $A(r)$ is different from the similar term in EYMH case due to the existence of a factor $\frac{1}{r^2}$. The diagonal Einstein equations give,
\begin{eqnarray}
  -g'\left(\frac{1}{r}+\frac{c'}{2c}\right)-g\left(\frac{1}{r^2}+\frac{3c'}{r c}+\frac{c''}{c}\right)+3
   &=& \frac{e^{\chi}}{4}A'^2+\frac{g}{4r^2c}B'^2+e^{\chi}\frac{q^2A^2B^2}{4r^2gc},  \nonumber\\
  -\frac{\chi'}{r}+\frac{c'}{c}\left(-\chi'+\frac{g'}{g}\right) &=& \frac{e^{\chi}q^2A^2B^2}{g^2r^2c^2}, \nonumber\\
  cc''+cc'\left(\frac{g'}{g}+\left(\frac{2}{r}-\frac{\chi'}{2}\right)\right) &=& -\frac{B'^2}{2r^2}+e^{\chi}\frac{q^2A^2B^2}{2g^2r^2}.
\label{maineq2}
\end{eqnarray}

The above equations are invariant under the following scaling symmetries:
\begin{eqnarray}
\label{rescale}
& &r \rightarrow a_1 r, \quad (t,x,y) \rightarrow (t,x,y)/a_1, \quad g \rightarrow a_1^2g, \quad A \rightarrow a_1 A, \quad B \rightarrow a_1 B, \\
\nn & &e^\chi \rightarrow a_2^2 e^\chi, \quad t \rightarrow a_2 t, \quad A \rightarrow A/a_2. \\
\nn & & x \rightarrow x/a_3,\quad B \rightarrow a_3B. \quad c \rightarrow a_3 c.
\end{eqnarray}
The second scaling symmetry may be used to set $\chi=0$ at infinity and the third scaling symmetry may be used to set $c=1$ at infinity, so that the asymptotic metric is that of $AdS_4$.

The fields have the following asymptotic behavior:
\begin{equation}
A = \mu - \frac{\rho}{r}, \quad B = B_0^b + \frac{B_1^b}{r},
\label{asymptotic}
\end{equation}
where $\mu$ is the chemical potential and $\rho$ is the charge density in the boundary theory. In what follows we will only consider the solutions for the field $B$ which vanishes near the boundary, i.e. $B_0 = 0$.

\section{Reissner-Nordstr\"{o}m black hole solution and Instability}
\label{sec:RN}
 To see when one expects hairy black holes at low temperature, one can study linearized perturbations of the extremal Reissner-Nordstr\"{o}m AdS (RN-AdS) black hole \cite{Horowitz:2009ij}. The general RN-AdS  solution is given by,
\begin{eqnarray}
c=1, \quad \chi=B=0, \quad
g=r^2-\frac{1}{r}\left(1+\rho^2/4\right)+\rho^2/4r^2, ~~~
A=\rho\left(1-1/r\right)\label{RN}
\end{eqnarray}
The temperature of the black hole (\ref{RN}) is
\begin{eqnarray}\label{temp}
 T=\frac{\left[g' (g~e^{-\chi})' \right]^{1/2}}{4\pi}|_{r=r_+}
\end{eqnarray}
For AdS-RN,  $T=(12-\rho^2)/16\pi$. It is known that fluctuations of $B_x$ in this background develops a tachyonic mode \cite{Gubser:2008wv} at low temperaure and for sufficiently large $q$. The resulting solution is a superconducting black hole with vector hairs. Following \cite{Horowitz:2009ij}, we argue that such an instability persists at the extremal limit. At the extremal limit ($T=0$), we have $\rho=2\sqrt{3}$. The near-horizon limit of the extremal solution is $AdS_2\times \mathbb{R}^2$ with a metric,
\begin{eqnarray}
ds^2=-6(r-1)^2 dt^2+\frac{dr^2}{6(r-1)^2}+dx^2+dy^2, ~~~
A=2\sqrt{3}(r-1)
\end{eqnarray}
Plugging this into the wave equation of $B(r)$,
and changing variables $\tilde{ r}=r-1$, we  recover a wave equation for
$AdS_2$ with a new effective mass,
\begin{eqnarray}
B_{,\tilde{ r} \tilde{r}}+\frac{2}{\tilde{ r}}B_{,\tilde{ r}}-\frac{m^2_{eff}}{\tilde{ r}^2}B=0,
~~~~~ ~m^2_{eff}=-\frac{q^2}{3}.
\end{eqnarray}
The instability to form $SU(2)$ vector hair at low temperature then is just the instability of scalar fields below the Breitenlohner-Freedman (BF) bound for $AdS_2$: $m^2_{BF} = -1/4$. Thus the condition for instability is
\begin{eqnarray}\label{unstable}
q^2>\frac{3}{4}.
\end{eqnarray}
It is reasonable to expect that such an instability takes the system to its superconducting ground state. 

\section{Zero temperature solution}
\label{sec:zero}

\begin{figure}
\begin{center}
\includegraphics[scale=0.7]{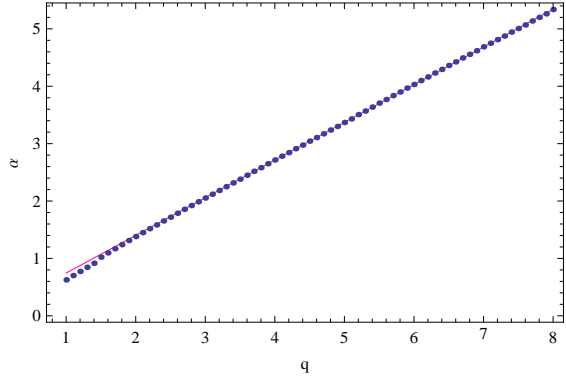}
\caption{$\alpha$ vs $q$ plot shows a almost linear relation with a slight dip for $\alpha$ for a value of $q<2.5$.  }
\label{fig:avsq}
\end{center}
\end{figure}

Being a single state without any degeneracy, a superconducting ground state does not have any entropy associated with it. This fact concur with the observations in \cite{Horowitz:2009ij},\cite{Gubser:2009cg,Gubser:2009gp} that the zero temperature limit of the superconducting black holes in EYMH theory have zero horizon size. Motivated by these facts we assume that same is true for a non-Abelian hairy black hole also and the ground state is a geometry with zero horizon size. We choose the following ansatz, near $r\to 0^+$, (similar to that of EYMH case \cite{Horowitz:2009ij})
\begin{eqnarray}
A\sim A_0(r) , ~~B\sim B_0-B_1(r) ,~~\chi\sim \chi_0-\chi_1(r),~~g\sim r^2+g_1(r),~~c\sim c_0+c_1(r) .
\label{ansatz}
\end{eqnarray}
All the terms with subscript $1$ are sub-leading and go to zero, faster than the leading part where it is applicable, as $r\rightarrow0$. Putting \eref{ansatz} in \eref{maineq}, we get find out the various terms in \eref{ansatz}. For example from the equation of motion for $A$,
\bea
r^2 (r^2 A_0')'=\frac{q^2 B_0^2} {A_0}
\Rightarrow A_0 \sim e^{(-\frac{\alpha}{r})}, \quad \alpha=q B_0/{c_0},
\eea
where we have used the observation $A_0 \rightarrow 0$ at the horizon. Following similar procedures, we get
\begin{eqnarray}
A\sim A_0 e^{-\alpha/r}, ~~~B\sim B_0\left(1-\frac{e^{\chi_0}q^2A_0^2}{4\alpha^2}e^{-2\alpha/r}\right),
~~~c\sim c_0\left(1+\frac{e^{\chi_0}A_0^2}{8r^2}e^{-2\alpha/r}\right), \\
\nn \chi\sim \chi_0-\frac{e^{\chi_0}A_0^2\alpha}{2r}e^{-2\alpha/r},~~~g\sim r^2-\frac{e^{\chi_0}A_0^2\alpha}{4r}e^{-2\alpha/r},
\label{scaling}
\end{eqnarray}
where by rescaling (\ref{rescale}) we set the coefficients $A_0=1$ and $\chi_0=0,c_0=1$. After we solve the equations we again use the rescalings of $g,c,\chi$ to make the Asymptotic geometry the same as that of $AdS_4$. For a given $q$, we choose $\alpha$ in such a fashion that $B$ vanishes near the boundary \cite{Horowitz:2009ij}. The numerical evaluation shows us an almost linear relation between $q$ and $\alpha$ (Fig \ref{fig:avsq}),with a slight dip for a lower value of $q$. Fitting a linear relation between $q$ and $\alpha$ with a range of $3<q<8$, we get $\alpha \approx 0.088 + 0.66 q$.

Our numerics is less reliable in the region $q<0.95$. It is not clear what exactly happens in this region. As $q \rightarrow \frac{\sqrt{3}}{2}\approx0.866$, the geometry of our solution seems to approaches extremal RN geometry (Fig \ref{fig:avsq}). However the questions of phase transitions remain unclear. It is expected that there is a second order phase transition at $q=\frac{\sqrt 3}{2}$ from RN black hole to our solution. However one may not rule out the possibility of a first or any other type of phase transition(s). At least in the range of our numerics it seems that our solutions have lower free energy than the extremal $RN$ black hole in a grand canonical ensemble. As expected the solitonic solution is more massive than the RN black hole, but they carry more charge for the same boundary chemical potential and consequently have a lower free energy.
\begin{figure}
\centering
\subfigure{
\includegraphics[height=1.8in]{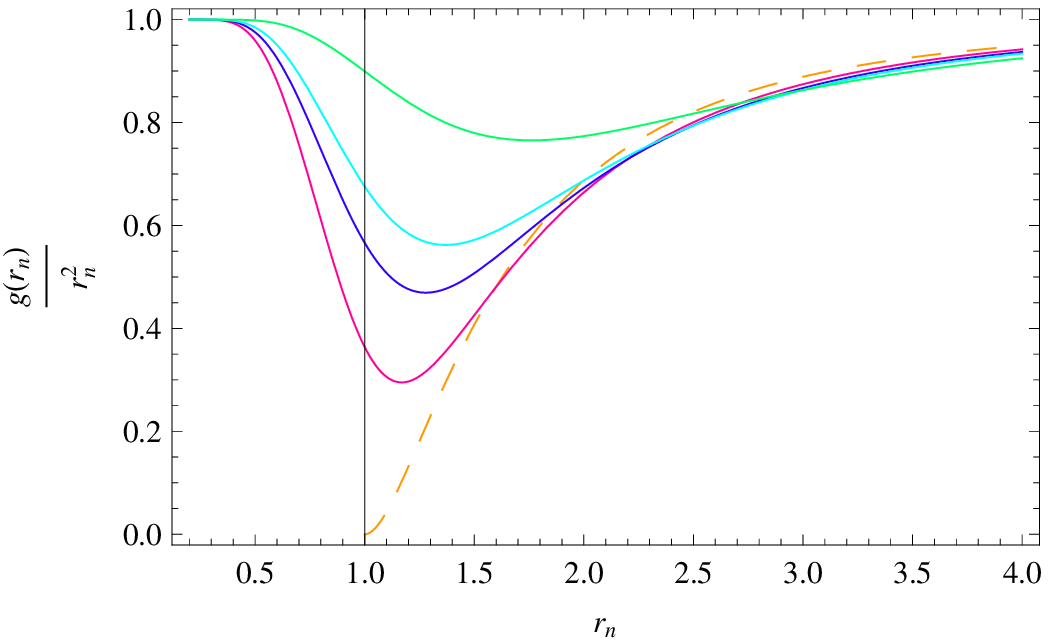}
\label{fig:gvsr}
}
 \subfigure{
 \includegraphics[height=1.8in]{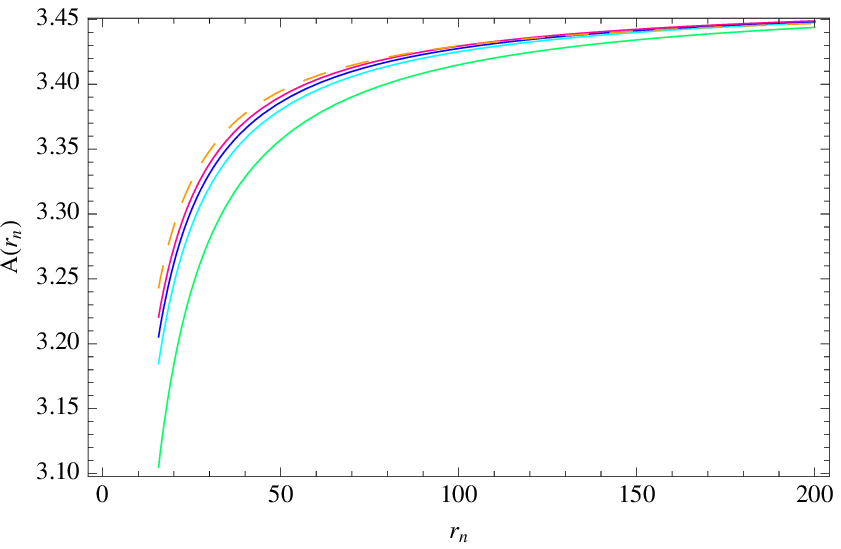}
 \label{fig:gvsr2}
 }
\caption{The left panel shows a plot of the rescaled $\frac{g}{r^2}$ with the rescaled variable  $r_n$. From top to bottom $q=1.5,1.1,1,0.9$. The dashed line is for the RN black hole. The right panel shows a plot of $A$ with $r_n$ for the same set of values of $q$, from bottom to top.}
\end{figure}

    General properties of our solution is similar to EYMH case \cite{Horowitz:2009ij}. $B,\chi$ approaches a constant near $r=0$, $g=r^2$ and the gauge field $A$ and its derivative (electric field) vanishes. The metric approaches $AdS_4$ with the same value of the cosmological constant at infinity. The extremal horizon is just the Poincare horizon of near horizon geometry $AdS_4$. In terms of the dual field theory, this means that the full conformal symmetry is restored in the infrared. Due to the asymmetry of our solution in $x$ and $y$ direction the ratio of speed of light in IR and UV is not only non-unity but also different in different directions. We have,
\begin{eqnarray}
\frac{v_x^{IR}}{v_x^{UV}}=\frac1{c_0} \exp(\frac{\chi_0}{2}), ~~\frac{v_y^{IR}}{v_y^{UV}}=\exp(\frac{\chi_0}{2})
\end{eqnarray}

   Our solution may be viewed as static, charged solitons, where the electrostatic repulsion is balancing the gravitational and non-Abelian interactions. As can be checked easily from the asymptotic behaviors the energy of the soliton is finite. \footnote{However it should be noted that $B(r)$ is not square integrable. This is not a requirement of a valid solution to the Einstein's equations and assuming that our numerics are reliable, such solutions exists. The question is that whether they play a role in the context of AdS/CFT. In a sense whether it belongs to same quantum ensemble as the extremal black hole. As our zero temperature solution is of finite energy this seems to be the case. There are well known solutions such as the BPST instanton \cite{Belavin:1975fg} which also have a non square integrable gauge field configuration with a bounded action. In our case, just as in case of BPST instanton, the solution may be approached by configurations which are of finite norm with a bounded energy (just replace $B \rightarrow \exp(-\frac{\beta}{r}) B(r)$, $\beta \rightarrow 0$). At finite temperature the gauge field $B$ is likely to be square integrable due to the cutoff provided by the horizon. Our zero temperature solution may also be thought as a zero temperature limiting case of such a solution (see Sec. \ref{sec:finite}).} One interesting feature is the existence of an ``essential singularity''(i.e. an $e^{-\frac{\#}{r}}$ behavior) of $A(r)$ at $r=0$. Similar singularities appear for other quantities in \eref{scaling}. Hence all the sufficiently high derivative of the metric and gauge fields vanishes at $r=0$.  This is in contrast with the EYMH case \cite{Horowitz:2009ij},   where sufficiently large curvature invariants diverge at $r=0$ for a generic  $\alpha$ and behavior of the fields near $r=0$ is power law. In terms of the ``tortoise'' co-ordinates (\ref{tortoise}), the essential singularity is just the reflection of the fact that $A$ is massive near the horizon and decays exponentially.

\subsection{Conductivity}
\label{sec:conductivity}

In order to calculate the conductivity of this system, we need to turn on a small perturbation in the vector potential. The gauge field and the associated metric perturbations are of the form:
\begin{equation}
\label{perturb}
A_y^3 = \epsilon a(r)e^{-i\omega t}\tau^3 dy, \qquad g_{ty}=\epsilon h(r)e^{-i\omega t},
\end{equation}
where we assume that both the gauge and the metric perturbations are of the same order $O(\epsilon)$.  We won't be looking in the other component of the gauge field perturbations which decouples from the above. From Maxwell's equation we get,
\begin{eqnarray}
A_y^3\longrightarrow a''+a'\left(\frac{g'}{g}-\frac{\chi'}{2}+\frac{c'}{c}\right)
  +a\left(\frac{e^{\chi}\omega^2}{g^2}-\frac{q^2B^2}{gr^2c^2}\right)
  +\frac{e^{\chi}h}{g}\left(A''+A'\left(\frac{h'}{h}+\frac{\chi'}{2}+\frac{c'}{c}\right)-\frac{q^2B^2}{r^2gc^2}A\right)=0. \nonumber\\
\end{eqnarray}
To linear order, Einstein's equations give,
\begin{eqnarray}
T_{ry}\longrightarrow -\frac{2}{r}h+h'=-a A',
\end{eqnarray}
which combined with the equation of $A(r)$ to give:
\begin{eqnarray}
a''+a'\left(\frac{g'}{g}-\frac{\chi'}{2}+\frac{c'}{c}\right)
  +a\left(\frac{e^{\chi}\omega^2}{g^2}-\frac{q^2B^2}{gr^2c^2} -e^{\chi}\frac{A'^2}{g}\right)=0.
\end{eqnarray}

\begin{figure}
\begin{center}
\includegraphics[scale=0.75]{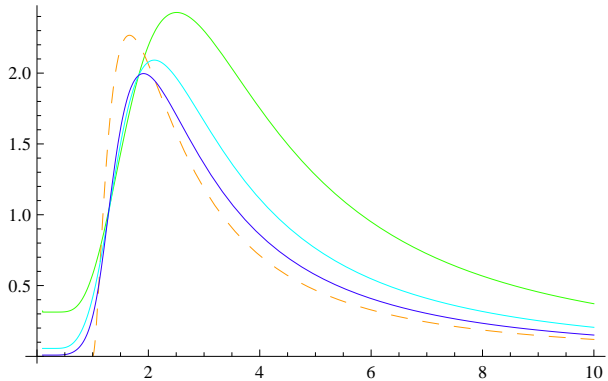}
\caption{Plot of the rescaled potential $V(r_n)$ for $q=1.5,1.1,1$ (from the right) at $T=0$ with rescaled co-ordinate $r_n$. The poincare horizon is at $r_n=0$ for the zero temperature solutions. $V(r_n)$ is non-zero at the horizon. The dashed curve is for the extremal black hole with a horizon at $r=1$. }
\label{fig:poplot}
\end{center}
\end{figure}

This can be written as a Schr\"{o}dinger equation:
\begin{equation}
 \label{schroedinger}
 -a'' + V(\tilde{r})a = c^2 \omega^2 a,
\end{equation}
where
\begin{eqnarray}
V(r)=g\left(c^2A'^2+e^{-\chi}\frac{q^2B^2}{r^2}\right).
\label{vr}
\end{eqnarray}
and all the derivatives in \eref{schroedinger} are in terms of the new variable new variable $\tilde{r}$ (``tortoise coordinate'') given by:
\begin{equation}
 \frac{d}{d\tilde{r}} \equiv e^{-\chi/2} gc \frac{d}{dr}.
\label{tortoise}
\end{equation}
Here, $c$  approaches to unity as $r \rightarrow \infty$, so that the spacetime is asymptotically $AdS_4$. It follows then from  \eref{asymptotic} that the potential $V(r)$ vanishes near the boundary. If we require $g\sim r^2$ near the horizon at $r=0$ then the first term vanishes, while the second term is finite as $B(r=0) \equiv B_0 \neq 0$ and $\chi$ is also finite at the horizon. The value of the potential at the horizon decreases as $q \to \frac{\sqrt{3}}{2}$ (Fig \ref{fig:poplot}). Note that since $c \rightarrow 1$ near the boundary, the quantity $\omega$ can be interpreted as the frequency of the incoming wave.

The superconducting nature of the system is argued from the existence of a supercurrent solution. If we choose $\omega=0$ and integrate $A^3_x$ from the horizon (with a regularity condition at the horizon), we are expected to get a non-trivial $A_x$. Existence of such a solution implies a $\delta$ function for the real part of conductivity at $\omega=0$ \cite{Basu:2008st,Horowitz:2009ij} \footnote{One should be more careful about the existence of a delta function for the real part of conductivity at $\omega=0$ and relation between superconductivity. Due to the coupling of gravity modes the potential $V(r)$ is non-trivial even in the  normal phase (Fig \ref{fig:poplot}) and consequently a delta function is present in the real part of DC conductivity. Physical interpretation of such a delta function comes from the ``translational invariance'' of the theory. See \cite{Hartnoll:2008kx} for further discussions.}
\begin{equation}
 \mathrm{Re}(\sigma(\omega)) \sim \delta(\omega)+\mathrm{finite}.
\end{equation}

The nature of the finite part of the conductivity can be inferred from the potential $V(r)$. As argued before the potential vanishes near the boundary and is finite near the horizon. The fact that the potential is nonzero at the horizon makes it possible for this system to exhibit a hard gap. From \eref{schroedinger}, the field $a$ has the following asymptotic behaviours near the horizon ($\tilde{r} \rightarrow -\infty$) and the boundary ($\tilde{r} \rightarrow 0$):
\begin{eqnarray}
 a(\tilde{r} \rightarrow 0) \sim a_0^b + a_1^b \tilde{r} \\
 a(\tilde{r} \rightarrow \infty) = a_0 e^{i \tilde{\omega} \tilde{r}},
\end{eqnarray}
where $\tilde{\omega} = \sqrt{c_0^2 \omega^2 - V_0}$, with $c_h$, $a_h$ being the near-horizon values of $c$ and $a$ respectively. Here, we have chosen the incoming boundary condition near the horizon. The conductivity of the system is given by:
\begin{equation}
 \label{conductivity}
 \sigma = -\frac{ia_1^b}{\omega a_0^b}
\end{equation}

It follows from \eref{schroedinger} that :
\begin{equation}
 a^* a'' - a a^{*''} = 0,
\end{equation}
which implies that the quantity $\Lambda = a^* a' - a a^{*'} = 2i \mathrm{Im}(a a^{*'})$ is a constant. Equating the values of $\Lambda$ near the horizon and the boundary we get:
\begin{equation}
 \mathrm{Re}(\sigma) = \left\{
\begin{array}{l l}
\frac{\tilde{\omega}}{\omega} \frac{|a_0|^2}{|a_0^b|^2} & , \quad \tilde \omega^2 > 0 \\
                         0,& \quad  \tilde \omega^2 < 0 \\
\end{array} \right.
\label{finalcond}
\end{equation}
Therefore, the real part of the conductivity will vanish whenever $\tilde{\omega}$ is imaginary, i.e. when $\omega < \sqrt{V_0}/c_0$, which defines the gap.

As our solution is an anisotropic superconductor (only superconducting in $A_y$ direction), the excitations in $A^3_x$ channel may not be gapped \cite{Gubser:2008wv}. The gap seems to be a property of the anisotropic non-Abelian system, as the scalar-Abelian gauge field system does not exhibit this gapped behaviour in the extremal limit \cite{Horowitz:2009ij}.

\section{Finite temperature analysis}
\label{sec:finite}
Although the \eref{maineq} has been solved in a probe limit, the full  set of equations with gravity backreaction (\eref{maineq2}) is not yet solved at a non-zero temperature. Hence it is an issue whether our zero temperature solution may be realized as a zero temperature limit of the finite temperature hairy black hole (and whether such a low temperature hairy black hole solution at all exists.). One affirmative clue comes from the emergent conformal symmetry in the IR. In the near horizon region, i.e. $\alpha \gg r$, our solution approaches a $AdS_4$ geometry with a background gauge field $A_x=B_0$ and $A_t=0$. One may consider a black hole situated deep inside this $AdS$ space, such that the radius of the black hole $r_0 \ll \alpha$. Here, $g(r)$ near the horizon is given by, $g(r)=r^2 (1-\frac{r_0}{r})$. However this Schwarzschild like solution will get correction due to a small non-zero $A(r)$ near the horizon. The solution of $A(r)$ in the $r\ll \alpha$ is given by,
\begin{eqnarray}
 A(r) \approx \exp(-\frac{\alpha}{r}) (1-\frac{r_0}{r}).
\end{eqnarray}
$A(r)$ is small in the region $r \ll \alpha$. In this region we can do a perturbation in $A(r)$ for other quantities. Realizing that our scaling relations \eref{scaling} is just a perturbation in $A(r)$, we can match a near horizon solution with the scaling solution (i.e. \eref{scaling}) in a region $r_0 \ll r \ll \alpha$. Hence we use the method of matched asymptotic expansion to argue that,
\begin{eqnarray}
  A_T(r)= A(r) (1-\frac{r_0}{r})\\
 \nn B_T(r)=B_0-(B(r)-B_0)(1-\frac{r_0}{r})^2 \\
 \nn c_T(r)=c_0+(c(r)-c_0)(1-\frac{r_0}{r})^2 \\
 \nn \chi_T(r)=\chi(r)+(\chi(r)-\chi_0)(1-\frac{r_0}{r}) \\
\nn g_T(r)=g(r)(1-\frac{r_0}{r})
\label{finiteT}
\end{eqnarray}
is a solution to Einstein's equations if $r_0 \ll \alpha$. The quantities ($A,B.c,\chi$) appear in the right hand side of the above equations are our numerical solution and the quantities with a subscript $T$ are the finite temperature versions. Whether correction to such a solution is finite or not is an open question, which may be partially settled by a convincing finite temperature numerics. However due to the possible ``smooth'' nature of the equations \eref{maineq}, i.e. a small change in the near horizon boundary condition leads to a small change near the boundary, it is reasonable to believe that corrections to \eref{finiteT} are well controlled.
\begin{figure}
\begin{center}
\includegraphics[scale=0.5]{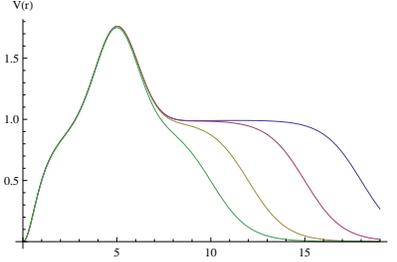}
\caption{Schematic plot of the potential $V(\tilde r)$ for different $T$. $T$ decreases from the left. $V(r)$ is gradually widened as $T$ decreased. Horizon is at $\tilde r=\infty$.}
\label{fig:widen}
\end{center}
\end{figure}
What happens to the potential $V(r)$ (see \eref{vr}) in this limit is interesting. At any non-zero $r_0$, the potential $V(r)$ goes to zero at the horizon ($r=r_0$) and the height of the potential is always finite. Hence the transmission through this potential is non-zero. In our language the (i.e. \eref{finalcond}) $\tilde \omega=\omega$ is a real number. Hence the real part of the conductivity is non-zero for any value of $\omega$. However, in the ``tortoise co-ordinates'' (\ref{tortoise}) the potential is gradually widened as $r_0 \rightarrow 0$ (Fig \ref{fig:widen}). Hence the transmission through the potential gradually decreases as the horizon size is decreased. We expect a $\exp(-\frac{\Delta}{T})$ fall off for the real part of conductivity. A detailed discussion of the various gaps etc is left for a future work.

\section{Acknowledgements}
We thank Matthew Roberts, Mark van Raamsdonk and Moshe Rozali for discussions. We would also like to thank the String Theory Group at UBC for their support and encouragement. PB and AM acknowledge support from the Natural Sciences and Engineering Research Council of Canada.

\bibliographystyle{unsrt}
\bibliography{nonab.bib}
\end{document}